# WHAT ARE THE CONTROLS REQUIREMENTS FOR THE GLOBAL ACCELERATOR NETWORK?

Reinhard Bacher, Philip Duval and Stephen Herb
Deutsches Elektronen-Synchrotron DESY, D-22603 Hamburg, Germany

Abstract

The Global Accelerator Network is a proposed model for remote operation of a future large accelerator by the partners of an international collaboration. The remote functionality would include not only routine operation and machine studies, but also hardware diagnostic tests and software maintenance and development, so the control system must offer both high and low level remote access to the accelerator and its components. We discuss in this paper some of the special requirements, and how they might be satisfied using current or developing technology, such as web-based applications and industrial automation standards. The choices made must be acceptable to the collaboration, but will also have a strong influence on its effectiveness. The biggest technical challenge is likely to be the middle layer software which is responsible for system integration, automation, and presentation of the machine state to the operating staff.

## 1 INTRODUCTION

The premise of the Global Accelerator Network (GAN) [1] is that a large accelerator facility would be operated remotely by collaborating institutions, with reduced dependence on on-site personnel; many astronomical observatories are already operating in this mode. It is also likely that the accelerator would be built collaboratively, with groups and institutions responsible for construction, commissioning, and maintainance of particular subsystems [2]. Much of the specialized work necessary for tuning and maintenance of the subsystems would then also be performed via remote access.

Remote (and collaborative) operation will place additional demands on the control system and its software. At the same time, the problems will have considerable overlap with those of large industrial automation projects, and of businesses with operations distributed over the internet. Our intention in this paper is consider the structure of a GAN control system, with respect to the needs of the users, and with respect to some current directions in software development.

## 2 THE USERS AND THEIR NEEDS

The control system will have many different users:
- Machine operations crews
- Machine coordinators
- Accelerator physicists
- The experimental groups
- Automated 'middle layer' processes
- Hardware experts responsible for subsystems.
- Software experts responsible for subsystems.
- Software experts responsible for system integration and functionality
- IT System Administrators
- Crews responsible for local maintenance of accelerator and utilities subsystems

Their needs will range from publicly accessible overviews to highly specialized expert interfaces, and from low level device access to single knob control of global machine parameters. Many users will want standardized displays, while others will need to integrate specialized 'third party' software into the control structure. It seems unlikely that everyone will be satisfied by a single control system with the ansatz 'one size fits all'; on the other hand, freedom of choice at all levels would be a recipe for disastrous fragmentation. This is the most important structural issue for the control system.

## 3 DEALING WITH DIVERSITY

We can therefore assume that we will have to accomodate a range of computer platforms and software interfaces, without drowning in the resulting complexity. Here are some suggestions:

### 3.1 Multiple Interfaces for Front End Systems

With modern operating systems and software layering, it is practical on a single computer to provide hardware access over several interfaces to satisfy the needs of different users. Figure 1 shows several possibilities ('RPC' is taken here to stand for the messaging system used for control system integration, and 'IP' for the Internet Protocol interface to the network).

*Fieldbus Gateways* (Fig. 1a) permit experts to perform low level diagnostic tests, knowing that the results are not confused by an intermediate 'black box' software layer. They must be used with caution, since low level access may bypass device safety code in the main software path.

*Web Servers* (Fig 1b) may be used for diagnostic access. The hardware maintenance personnel often have needs very different from those of machine operation, and the web server interface may permit them to perform much of their own software development and maintenance work.

*LabView* is taken as an example of the high level graphical programming tools available for hardware control and integration. Many engineers now use these tools to develop sophisticated control and data acquisition systems; because of the impressive level of commercial support for the device drivers, it is often possible to avoid writing low level code, and this can result in large gains in convenience and productivity. Fig. 1c shows the case that the LabView application is used as an alternative diagnostic interface. It is also possible to connect the LabView applications to the 'RPC' interface, as in Fig 1d, using 'C' library calls or, in the Microsoft world, ActiveX components, so that the logic prepared by the engineer need not be duplicated. We have considerable experience with interfacing LabView and HPVee applications, and both require, for effective integration into the control system, a very serious support effort and close coordination with the engineer. The bottom line is that each such tool must either be well supported, or limited to a purely diagnostic role.

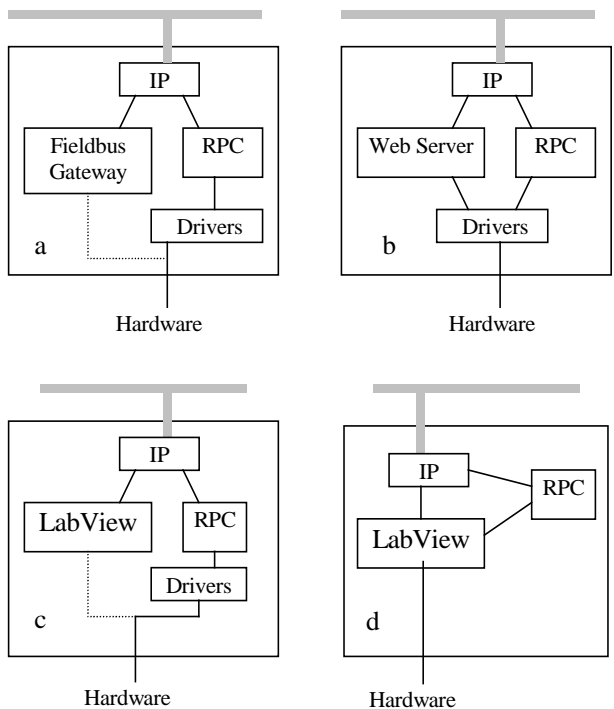

Figure 1: see text

### 3.2 Gateways Versus Software Busses

One means of integrating heterogeneous control subsystems is to introduce intermediate gateways for protocol translation. This turns out in practice to be extremely problematic; there are usually mismatches which cannot be bridged by reformatting of single data packets. It can be especially devastating for automation software which is attempting to coordinate the operation of underlying subsystems; if for example the error reporting from each subsystem is differently organized, responses must be hand-coded, rather than (at least partly) generic. The result is that the use of gateways can greatly complicate system integration.

It seems to be generally agreed that "Software Busses" are a better solution, which raises the question of whether they are more than a collection of gateways? Our limited understanding of this is that a software bus should enforce object models for representation of data and devices, and that each underlying subsystem must prepare an interface which conforms to the models. This has several consequences:
- translation takes place at the subsystem level, rather than the network protocol level, and the subsystem developer, who understands his system, is responsible for the interface
- if the software bus models are well defined in advance, the subsystem developer may partly incorporate them in his underlying code.

In fact it is desirable that the bus object models also be adapted to the subsystems during development, since object representations benefit from contact with reality.

Until recently, software busses have been more widely discussed than implemented, but integration needs in the business world are now driving a rapid development of standards and commercial tools, some of which will coincide with our requirements [3]. If the bus has adequate real time performance, it would be possible to also use it in some of the subsystems for internal messaging, obviating the need for multiple controls interfaces. This is a strong motivation to make the bus, and associated tools, as attractive as possible to the people developing these systems.

### 3.3 Specification And Emulation

A major problem with software which interfaces to hardware devices is that realistic tests are not possible until the hardware is available. The problem is even more severe at the system integration level, where the software is trying to coordinate the activities of many underlying subsystems; the higher layers of software cannot be tested until the lower levels (with hardware) are in place. For a large collaborative project, this is not acceptable. Obviously part of the solution lies in formal specification procedures and modelling, but an equally important ingredient is hardware (and in some cases software) emulation, which permits system integration tests to start before all the hardware is in place. Perhaps it should be required that each group responsible for a particular subsystem should supply an emulation program which implements at least some fraction of the specifications and can be used by collaborators as a 'plug-in' substitute for the final product. Our experience is that even a modest effort in this direction is extraordinarily useful, also after commissioning, since software updates can be tested without disturbing accelerator operation.

*3.4 Multiple Interfaces at the Console Level*

Many of the user interface requirements may be satisfied using 'Web service' techniques, which will continue to improve in scope and sophistication. But the accelerator turn-on and R&D work will involve non-standard operation with special measurement sequences and calculations, making use of a mixture of automated sequences and low level access to the hardware. Tools such as MathLab permit accelerator physicists to script complicated sequences and calculations with binding via 'C' function calls to the control system API, and could provide alternative console level interfaces to the control system. As with LabView at the front end, each such tool requires a significant support effort for effective integration with the controls.

A related point is whether the console software should run on the remote console computers, (so that the messaging extends to the remote systems) or be exported from the local computers by techniques such as X-windows or Web pages. There are arguments in favor of both, and a mix would be possible.

*3.5 Configuration Wizards*

RAD tools are often effective within specialized niches, and many of the jobs for interfacing front end and console applications to the software bus may be candidates for 'Wizards' which guide the non-expert user through a setup which produces a code skeleton (or even a finished application) for his chosen platform. This can be an efficient means of collaboration between hardware engineers and the software group.

There is currently much enthusiasm (for example [4]) for maintaining GUI applications in XML and "rendering" the XML to the desired platform. Such a strategy could have great benefits at the console level, where different views of the same application could produce versions for a web browser, or a WAP handy.

# 4 DIRECTIONS IN INDUSTRIAL CONTROLS

In the past, control systems for industrial automation were based on (more or less) proprietary chains of hardware and software. This is now changing, and many of the new developments are closer to our controls needs, and to our natural preference for internet technologies.

*4.1 'OPC' and 'OPC DX'*

In the past decade, industrial PCs running Microsoft DOS and Windows have become an important part of the controls and automation environment; tools such as OLE and ActiveX have been eagerly adopted and have provided great benefits in cost, ease of use, and connectivity with other software tools. For communications, 'OPC' ("Ole for Process Control") is a widely used standard for access to process data. A new standard under development, 'OPC DX' [5] appears to be, in effect, a software bus providing Ethernet connectivity between heterogeneous fieldbus clusters and overlying monitoring systems. Work is also underway to incorporate features of the Microsoft .NET environment. At present this is limited to PCs running Microsoft Windows, but Wind River has recently announced support for OPC under VxWorks on a range of architectures [6]. We must assume that we will need the ability to communicate with OPC-based systems. An interesting question is whether there will associated management tools which might make it profitable to adopt some of the standards.

*4.2 Ethernet in Automation*

The need for greater flexibility in industry and the demonstration by the internet of the immense benefits of interoperability have led in the past several years to great interest in moving industrial controls toward IP-based communications with front-end hardware. The suitability of Ethernet for replacing traditional fieldbusses is a complicated and controversial subject, but several industry groups are now trying to develop standards, including object descriptions which will permit integration of heterogeneous front-end devices in automation procedures [7].

A related aspect is the adoption by industry of the Internet Appliance model, whereby a device or machine connects directly to an IP network, permitting, for example http access to documentation and configuration or diagnostic procedures. Much of the development is in the direction of small embedded processors (often running Linux variants), and the cost of such interfaces is falling rapidly. This is obviously well suited to our remote access needs.